\documentclass[a4paper,10pt,pra,twocolumn]{revtex4}
\usepackage{graphicx}
\usepackage{amssymb,amsthm,amsfonts,amstext}
\usepackage{url}
\def\be{\begin{equation}}
\def\ee{\end{equation}}
\def\bea{\begin{eqnarray}}
\def\eea{\end{eqnarray}}
\def\bma{\begin{mathletters}}
\def\ema{\end{mathletters}}
\def\P{{\cal P}}
\def\0{\overline{0}}

\def\q0{\underline{0}}

\def\H{{\cal H}}

\def\id{{\mathbb I}}
\def\E{{\cal E}}

\def\H{{\cal H}}

\def\tr{\mbox{tr}}
\def\one{\leavevmode\hbox{\small1\normalsize\kern-.33em1}}

\newtheorem{theo}{Theorem}

\newtheorem{lemma}[theo]{Lemma}

\def\id{{\mathbb I}}
\def\tr{\mbox{tr}}
\begin{document}

\title{A glance beyond the quantum model}

\author{Miguel Navascu\'es$^{1,2}$ and
        Harald Wunderlich$^{1,2,3}$}
\affiliation{$^1$Institute for Mathematical Sciences, Imperial College London, SW7 2PE, UK,\\
$^2$QOLS, The Blackett Laboratory, Imperial College London,
 Prince Consort Rd., SW7 2BW, UK, \\$^3$Fachbereich Physik, Universit\"{a}t Siegen, 57068 Siegen, Germany}

\date{\today}

\maketitle

\textbf{One of the most important problems in Physics is how to reconcile Quantum Mechanics with General Relativity. Some authors have suggested that this may be realized at the expense of having to drop the quantum formalism in favor of a more general theory. However, as the experiments we can perform nowadays are far away from the range of energies where we may expect to observe non-quantum effects, it is difficult to theorize at this respect. Here we propose a fundamental axiom that we believe any reasonable post-quantum theory should satisfy, namely, that such a theory should recover classical physics in the macroscopic limit. We use this principle, together with the impossibility of instantaneous communication, to characterize the set of correlations that can arise between two distant observers. Although several quantum limits are recovered, our results suggest that quantum mechanics could be falsified by a Bell-type experiment if both observers have a sufficient number of detectors.}

At the beginning of the 21$^{st}$ century, one of the main goals of theoretical Physics is to come up with a theory that reconciles Quantum Mechanics and General Relativity. Currently, there are several approaches in this direction, like String Theory \cite{polchinski} or Loop Quantum Gravity \cite{rovelli}. What most of these approaches have in common is that they take the mathematical structure of Quantum Mechanics for granted and then try to find a suitable dynamics such that the resulting theory approaches General Relativity in some limit. The problem at stake is, thus, how to ``quantize gravity''. Such an approach may be condemned to fail, since it could very well be that Quantum Mechanics is not a fundamental theory of Nature, but an effective model, only valid within a specific range of energies. Indeed, some considerations about black hole evaporation suggest that certain axioms of Quantum Theory should be reexamined \cite{page}.

However, as the particle experiments we can perform nowadays are energetically very far from the Planck mass $m_P\approx 1.2 \times 10^{19}$ GeV/$c^2$, trying to formulate a new candidate beyond the quantum theory would be premature (although cosmological observations could also provide some insight). To solve the issue, different authors have suggested different alternatives. One approach, followed by Popescu and others (see, for instance, \cite{popescu,acin,barrett}), is to derive general results that should apply to any physical theory fulfilling a set of reasonable axioms.

It is agreed by most that the next candidate theory should prevent instantaneous communication between distant parties. If we take this as an axiom (the no-signaling principle), we can derive a set of restrictions to be satisfied by any physical theory compatible with it. A lot of properties of these theories are known: any nonlocal theory inside that set does not allow to replicate an unknown state \cite{barrett}, and it has to respect a specific set of uncertainty relations \cite{acin}. If the theories are non local enough (in particular, quantum theory), we can even have secret communication between distant parties \cite{masanes}.

Nevertheless, it seems that the no-signaling principle alone is not enough to prevent the existence of very ``unphysical'' theories. For that reason, several authors have considered that we could further restrict the set of allowed theories by adding more reasonable axioms to the no-signaling principle. That way, we have partial characterizations of the set of correlations allowed between distant parties in no-signaling worlds where the communicational complexity is not trivial \cite{brassard}, or where the efficiency of random access coding is limited by the number of bits we are allowed to communicate classically \cite{scarani,scarani2}.

In this paper, we propose a new axiom we call \emph{macroscopic locality}. The idea behind this axiom is that any physical theory should recover classical physics in the continuum limit, (i.e., when a large number of particles are involved and our measurement devices fail to resolve discrete particles). We will show that this very intuitive axiom, together with the no-signaling principle, allows to recover many important results in quantum mechanics, like the universality of the Tsirelson bound, or the set of accessible two-point correlators. Moreover, we will provide a complete characterization of the correlations between two distant physical systems that arise out of both axioms and comment on its consistency.

Although the set of possible correlations that can emerge out of these two principles is very similar to the quantum set, it is not identical. If we accept macroscopic locality as a fundamental law, this implies that a deviation from Quantum Mechanics could in principle be detected via a Bell-type experiment.

\section{Microscopic and Macroscopic experiments}


Suppose that we have two space-like separated parties, say Alice and Bob, in regions ${\cal A}$ and ${\cal B}$. In a \emph{microscopic} experiment of nonlocality (see Figure 1), there is a \emph{microscopic event} in some intermediate region that produces a pair of particles. One of the particles of the pair ends up in region ${\cal A}$, while the other one is received in region $B$. The moment the particles arrive at regions ${\cal A}$ and ${\cal B}$, Alice and Bob will subject them to an interaction. And the nature of this interaction is going to determine with which probabilities the particles will collide with the detectors in Alice's and Bob's labs.

We will identify the \emph{measurement settings} of Alice and Bob with the particular interactions $X$ and $Y$ they subject their corresponding particles. If Alice and Bob can each perform one out of $s$ possible interactions, the measurement settings of Alice will be numbered from 1 to $s$, while Bob's ones will go from $s+1$ to $2s$. 

The particular detectors that click after the experiment will determine the \emph{measurement outcomes}. To avoid a hypercomplicated notation, we will associate each measurement outcome $c$ to a pair of symbols $(Z,D)$, corresponding to the interaction $Z$ performed during the measurement procedure and the detector $D$ that received the particle. Outcomes corresponding to Alice's (Bob's) detectors will be denoted by $a$ ($b$), and will belong to the set $A$ ($B$) of Alice's (Bob's) possible outcomes. The application $X(a)$ ($Y(b)$) will return the measurement setting $X$ ($Y$) associated with that particular outcome, and $D(c)$ will label the physical detector related to the outcome $c$.

\begin{figure}
  \centering
  \includegraphics[width=9 cm]{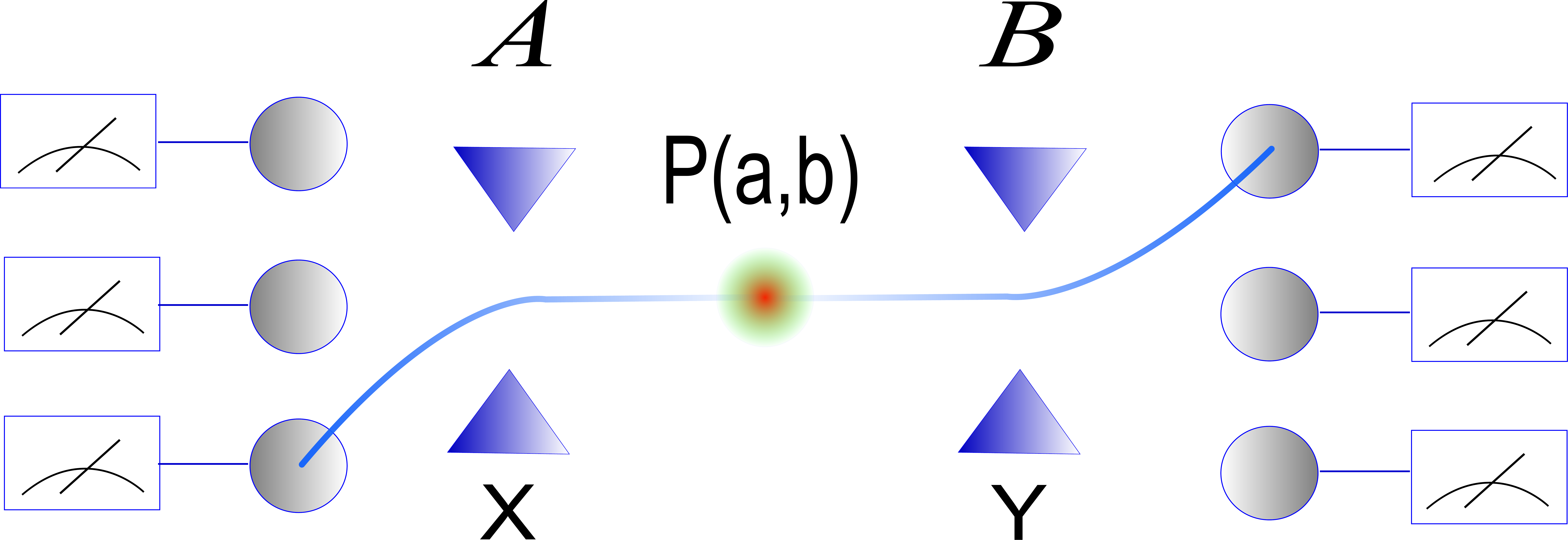}
  \caption{{\bf Microscopic experiment.} If Alice and Bob repeat an experiment many times, each time applying random interactions $X$ and $Y$, they can compare their measurement outcomes, make statistics and obtain a set of probabilities $P(a,b)$, the probability of detecting clicks in detectors $D(a), D(b)$ when Alice applies an interaction $X(a)$ and Bob applies an interaction $Y(b)$. From now on, we will assume that the no-signaling condition holds, i.e., that, for any $X,X'$ with $X\not=X'$, $\sum_{a\in X}P(a,b)=\sum_{a\in X'}P(a,b)\equiv P(b)$, and, for any $Y,Y'$ with $Y\not=Y'$, $\sum_{b\in Y}P(a,b)=\sum_{b\in Y'}P(a,b)\equiv P(a)$. These two conditions just assert that Alice's choice of measurement setting cannot affect Bob's statistics and viceversa. Also, the set of marginal probability distributions $P(a,b)$ in general will not admit a \emph{local hidden variable model}. This means that in some cases there will \emph{not} exist a joint probability distribution for all $2s$ possible measurements $P(c_1,...,c_{2s})$ such that $P(a,b)=\sum_{c} P(...,c_{X-1},a,c_{X+1},...,c_{Y-1},b,c_{Y+1},...)$.}
  \label{micros}
\end{figure}

A microscopic experiment is completely characterized by the set of probabilities $P(a,b)$ that Alice and Bob can estimate through statistical inference (see Figure \ref{micros}).

\begin{figure}
  \centering
  \includegraphics[width=9 cm]{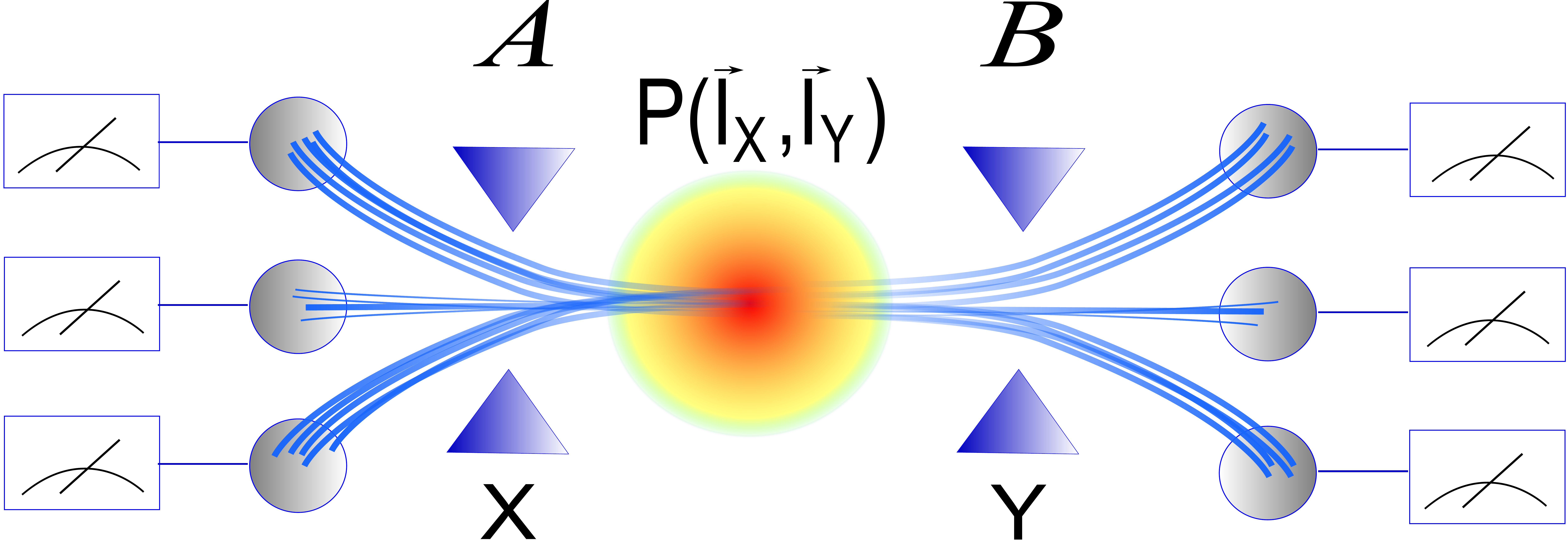}
  \caption{{\bf Macroscopic experiment.} Alice and Bob receive two particle beams, interact with them collectively and then take a record of the intensities measured on each detector. A macroscopic experiment implies restrictions on the physical states to be measured ($N$ identical and independent copies of a microscopic state), on the possible interactions to be performed (identical microscopic interactions over all the particles of each beam) and on the resolution of the detectors used (able to measure intensity fluctuations of order $\sqrt{N}$, but unable to resolve individual particles).
}
  \label{macros}
\end{figure}

On the other hand, in a \emph{macroscopic} experiment, a \emph{macroscopic event} will produce, not a pair of particles, but $N$ of them, with $N\gg 1$. This time, therefore, Alice (Bob) will not receive a single particle at a time, but a \emph{beam} of them. As before, Alice and Bob will be able to interact with these particles. However, they will not be able to address them individually, so, whatever microscopic interaction they intend to use, it will be applied to \emph{all} the particles of the beam at the same time. As a consequence of those interactions, the initial beam will be divided into several beams of different intensity that will collide with Alice's (Bob's) detectors, as shown in Figure \ref{macros}; a similar scenario was proposed in \cite{brunner3}.


Now we cannot associate measurement outcomes with clicks on a detector, since all detectors will click in each instance of the experiment. Thus in this scenario the measurement outcomes will be the distribution of intensities measured in each experiment. We will assume that the resolution of Alice and Bob's detectors does not allow them to resolve individual particles (i.e., we will be considering the \emph{continuum limit}). If the precision of their detectors is not very good, Alice and Bob will always observe the same distribution of intensities, the intensity registered at detector $D(a)$ being equal to $NP(a)$. However, if Alice and Bob have a resolution that allows them to measure changes in intensity values of the order $\sqrt{N}$, each time they repeat the experiment, they will observe fluctuations around this mean value $NP(a)$. 

\section{Macroscopic locality}

Under these conditions, we will say Alice and Bob's shared system exhibits \emph{macroscopic locality} if the distribution of intensities they can observe admits a local hidden variable model. More explicitly, if we denote the intensities measured by Alice's (Bob's) $d$ detectors by $\vec{I}_X$ ($\vec{I}_Y$), then Alice and Bob, through classical communication, will be able to estimate the marginal probability densities $P(\vec{I}_X,\vec{I}_Y)d\vec{I}_X d\vec{I}_Y$ for each pair of measurements $X,Y$. Their system will exhibit macroscopic locality iff there exists a \emph{global} probability density $P(\vec{I}_1,\vec{I}_2,...,\vec{I}_{2s})$ that admits $P(\vec{I}_X,\vec{I}_Y)d\vec{I}_Xd\vec{I}_Y$ as marginals. That is,

\be
P(\vec{I}_X,\vec{I}_Y)=\int\left(\prod_{Z\not=X,Y}d\vec{I}_Z\right)P(\vec{I}_1,\vec{I}_2,...,\vec{I}_{2s}).
\ee

In a sense, by imposing macroscopic locality, we are demanding that the distribution of intensities observed during a macroscopic experiment can be explained by a classical model. It is straightforward to generalize the notion of bipartite macroscopic locality to the multipartite case, i.e., to the case where more than two separate observers participate in an experiment. It is also clear that all classical (local) microscopic correlations remain local when driven to the thermodynamical limit.

\section{Characterization of macroscopic locality}

In this section we will show a simple characterization of the set of microscopic correlations that give rise to local macroscopic correlations. 

Our starting point will be to find a correspondence between the original microscopic distributions $P(a,b)$ and the final macroscopic probability densities we may observe when we bring the experiment to a higher scale. 

Suppose now that Alice and Bob switch their interactions to $X$ and $Y$, respectively, and define the microscopic observables $d_i^a$ ($d_i^b$) as equal to 1 when one of the particles of the pair $i$ impinges on Alice's (Bob's) detector $D(a)$ ($D(b)$) and 0 otherwise. It is clear that the intensities measured by Alice or Bob in the detector $D(c)$ will be proportional to $\sum_{i=1}^N d_i^{c}$. Here we are interested in the deviations of these intensities from their mean value, so, for simplicity, we will redefine these microscopic variables as

\be
\bar{d}_i^c=d_i^c-P(c).
\ee

\noindent That way, $\langle \bar{d}_i^a\rangle=\langle \bar{d}_i^b\rangle=0$, for all $X(a)=X,Y(b)=Y$. If we assume that Alice and Bob are able to measure fluctuations of the order $\sqrt{N}$, it makes sense to normalize these fluctuations as

\be
\bar{I}^{c}=\sum_{i=1}^N\frac{\bar{d}^{c}_i}{\sqrt{N}}.
\label{renorm}
\ee

According to the central limit theorem \cite{central}, for $N\to\infty$, the probability distribution governing the fluctuations of the intensities $\bar{I}^c$s will converge to a multivariate gaussian distribution, with $\vec{0}$ mean and covariance matrix $\gamma^{XY}$ given by $\gamma^{XY}_{cc'}=\langle \bar{I}^c \bar{I}^{c'}\rangle=\frac{1}{N}\sum_{i,j=1}^N\langle \bar{d}_i^{c} \bar{d}_j^{c'}\rangle=\langle \bar{d}_1^{c} \bar{d}_1^{c'}\rangle$, where in the last step we used the fact that different pairs of particles are uncorrelated. $\langle \bar{d}_1^c \bar{d}_1^{c'}\rangle$, in turn, can be easily expressed in terms of $P(a,b)$.

We have then that any set of microscopic correlations is mapped to marginal gaussian probability distributions. What we want to know is if this set of gaussian marginal probability distributions derives from a common probability distribution.

It is evident that, if such a global distribution exists, it should admit a positive semidefinite \emph{global} covariance matrix $\Gamma$, with $\Gamma_{cc'}=\langle\bar{I}^c\bar{I}^{c'}\rangle$, for all $c,c'\in A\cup B$. That is, there must exist a matrix $\Gamma\geq0$ of the form

\be
\Gamma=\left(\begin{array}{cc}Q&P\\P^T&R\end{array}\right),
\label{total}
\ee

\noindent where the columns and rows of $Q$ ($R$) are labeled by the elements of $A$ ($B$) and $P$'s rows are specified by elements of $A$ and $P$'s columns are specified by elements of $B$. The form of $P$ is completely determined, and given by

\be
P_{ab}=P(a,b)-P(a)P(b),
\label{correAB}
\ee

\noindent whereas $Q$ and $R$ are only partially determined by the relations

\begin{eqnarray}
Q_{a,a'}&=&\delta_{aa'}P(a)-P(a)P(a'),\mbox{ if } X(a)=X(a'),\nonumber\\
R_{b,b'}&=&\delta_{bb'}P(b)-P(b)P(b'),\mbox{ if } Y(b)=Y(b').
\label{correAA}
\end{eqnarray}

\noindent This way, the matrices $\gamma^{XY}$ defined above are submatrices of $\Gamma$. However, there are several entries of $\Gamma$ that are not known, like, for instance, $Q_{aa'}$, for $X(a)\not=X(a')$. Those would correspond to the correlations between intensity fluctuations corresponding to different settings, and, of course, we cannot measure them directly. However, if there exists a global distribution compatible with the previous gaussian marginals, then there must exist a set of real numbers $\{Q_{aa'},R_{bb'}:X(a)\not=X(a'),Y(b)\not=Y(b')\}$ such that the overall matrix $\Gamma$ is positive semidefinite. The inverse relation also holds, for if such numbers indeed exist, then there exists a global probability distribution from which all intensities arise, namely, the gaussian probability distribution with covariance matrix $\Gamma$ and $\vec{0}$ displacement vector. Therefore, we end up with a set of necessary and sufficient conditions that a set of microscopic correlations $P(a,b)$ has to satisfy in order to generate local macroscopic distributions. 

As shown in the Appendix \ref{charac_proof}, it can be proven that the set of microscopic correlations characterized above actually corresponds to the set of correlations $Q^1$, introduced in \cite{quantum} as a first approximation to the set $Q$ of quantum correlations. We will therefore call it $Q^1$ along the rest of the article. The set $Q^1$ admits a very simple numerical characterization via semidefinite programming \cite{quantum, sdp}.

\section{Quantum correlations are macroscopically local}

Next we will prove that $Q\subset Q^1$, that is, that all quantum correlations are compatible with macroscopic locality.

Suppose that our macroscopic (gaussian) correlations arise from a quantum mechanical system in a state $\rho=\sigma^{\otimes N}$, where $\sigma$ is the microscopic quantum state describing a single pair of particles. Then, following equation (\ref{renorm}), we can assign a hermitian linear operator $\tilde{I}^c$ to each of the intensity fluctuations, as measured in detector $D(c)$. Defining $\Gamma_{cc'}\equiv\tr(\rho\{\tilde{I}^c,\tilde{I}^{c'}\}_+)/2$ (where $\{\}_+$ denotes the anticommutator), we arrive at a real matrix of the form (\ref{total}), with $P,Q,R$ satisfying (\ref{correAB}), (\ref{correAA}). We will prove that such matrix is positive semidefinite. Let $\vec{v}$ be an arbitrary real vector. Then,

\begin{eqnarray}
\vec{v}^T\Gamma\vec{v}=& &\sum_{c,c'}v_c\Gamma_{cc'}v_{c'}=\tr(\rho (\sum v_c\tilde{I}^c)(\sum v_{c'}\tilde{I}^{c'}))=\nonumber\\
& &=\tr(\rho MM^\dagger)\geq 0,
\end{eqnarray}

\noindent where $M=M^\dagger=\sum v_c\tilde{I}^c$. Since $\vec{v}$ was arbitrary, this means that $\Gamma$ is positive semidefinite and, therefore, all quantum-mechanical systems are macroscopically local. It is easy to extend this result to prove that quantum correlations are also local in the multipartite case. However, the characterization of all possible microscopic correlations (quantum or not) compatible with nonlocality in the case of more than two observers seems a bit more complicated. For instance, in the case of three parties, we would not only have to impose macroscopic locality over the tripartite correlations $P(a,b,c)$, but also over the \emph{conditional} bipartite microscopic correlations $P(a,b|c)$ that would result if party ${\cal C}$ announced its measurement outcome.

\section{Closure under wiring}

\begin{figure}
  \centering
  \includegraphics[width=9 cm]{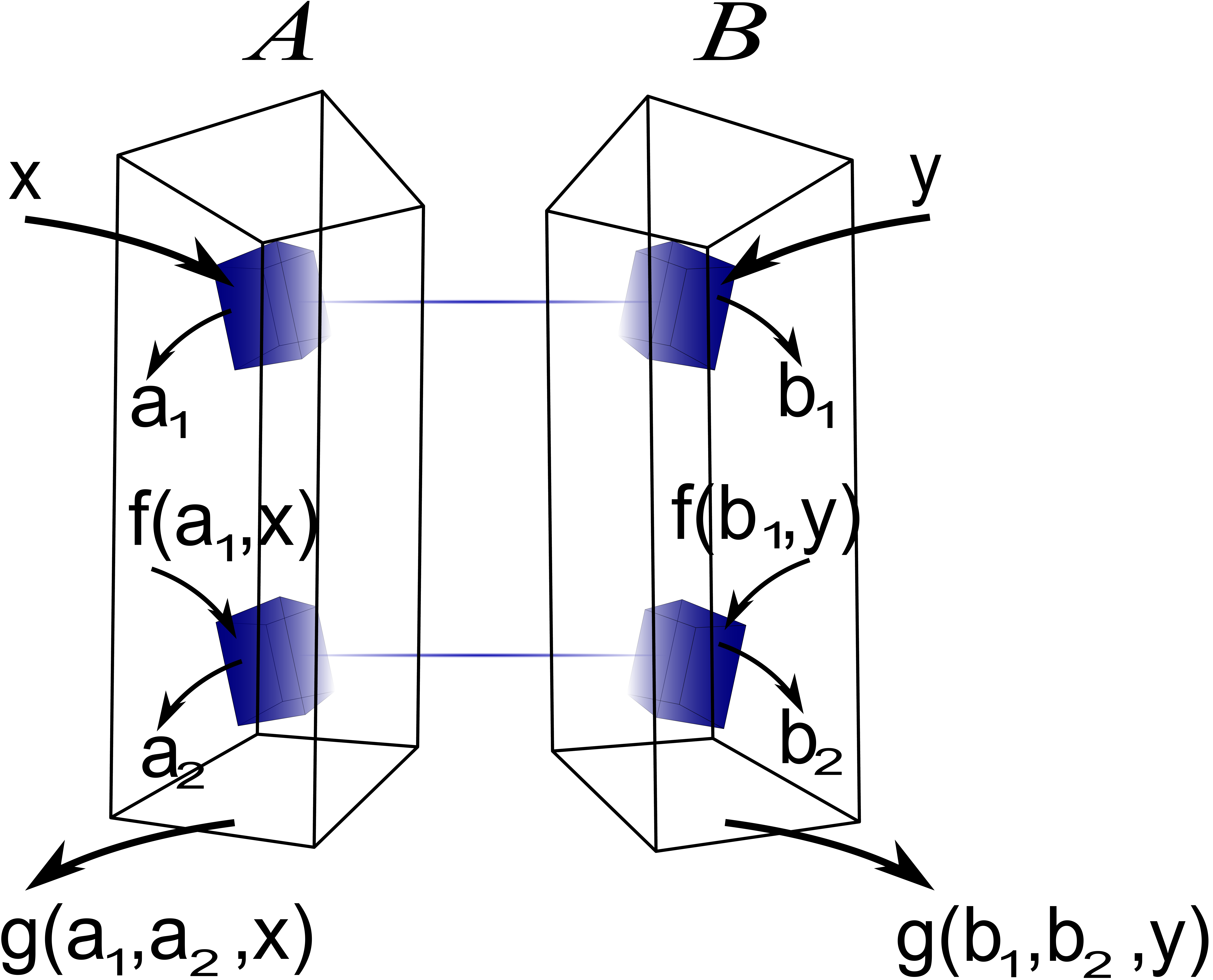}
  \caption{{\bf Wiring.} In \cite{brunner}, the authors showed that there exist some sets of correlations $P(a,b)$ very close to local from which extremely nonlocal correlations can be derived, provided that both parties have access to many copies of the given physical system. The key is to generate a new \emph{effective} set of correlations by measuring the $n$ systems sequencially, each time applying a measurement setting dependent on the (local) outcomes of the previously measured subsystems and our effective (local) measurement setting. Once all subsystems have been measured, the effective outcome of our virtual subsystem is then taken to be a function of all (local) measurement results. This process is known as \emph{wiring}. Above, you can see an example of deterministic wiring of two physical systems: the effective measurement $X$ ($Y$) of Alice's (Bob's) is applied over the first system, giving an outcome $a_1$ ($b_1$), while the interaction to be applied over the second subsystem is a function of both $X$ and $a_1$ ($Y$ and $b_1$). Labeling the second outcome by $a_2$ ($b_2$), the effective outcome of the whole scheme is a function of $a_1,a_2,X$ ($b_1,b_2,Y$). Note that, by definition, wiring is local, i.e., it does not require communication between Alice and Bob.}
  \label{wiring}
\end{figure}

The mechanism of using a set of independently correlated systems to generate a new single effective system of microscopic correlations (by conditioning the measurement settings of some systems on the measurement outcomes of some others) is called \emph{wiring} (see Figure \ref{wiring}), and we will denote it by ${\cal W}$. Since it can be used to increase locality violations \cite{brunner}, wiring poses a problem to our previous derivation of the set of correlations compatible with macroscopic locality. In principle, it could be possible that Alice and Bob shared two systems of microscopic correlations $P(a,b),Q(a,b)\in Q^1$, that, through some clever wiring ${\cal W}$, allowed to generate a new set of correlations $R(a,b)={\cal W}(P,Q)\not\in Q^1$. Any theory compatible with macroscopic locality could not therefore admit both sets of correlations, and a detailed classification of such theories would be very complicated. Or worse, it could also happen that, even though $P(a,b)\in Q^1$, ${\cal W}(P^{\otimes n})(a,b)\not\in Q^1$. $P(a,b)$ should thus not appear in any consistent theory compatible with macroscopic locality, and we would have to reconsider our definition of $Q^1$. This possibility is ruled out by the next result:

\vspace{10pt}

The set $Q^1$ is closed under wiring. That is: let $\{P_i\}_{i=1}^n$ be any set of $n$ microscopic behaviors, such that $P_i\in Q^1$ for all $i=1,...,n$, and let ${\cal W}(P_1,P_2,...)$ denote the effective set of correlations that results after some wiring of such behaviors. Then, ${\cal W}(P_1,P_2,...)\in Q^1$.

\vspace{10pt}

\noindent For a proof, see Appendix \ref{wiring_proof}.

\section{The limits of macroscopic locality}


Given a measurement $Z$, we can always define an \emph{observable} $O_Z$ by associating each possible measurement outcome $c$ of such measurement with a real number $O_Z(c)$. Given a measurement $X$ by Alice and a measurement $Y$ by Bob, the \emph{two-point correlator} of $O_X,O_Y$ is given by

\be
E_{XY}\equiv\langle O_XO_Y\rangle=\sum_{a,b\in X, Y}P(a,b)O_X(a)O_Y(b).
\ee

Suppose that we are in a scenario where $k=d=2$ (with the measurement settings ordered as $\overbrace{12}^A\overbrace{34}^B$), and consider the set of observables $\{O_Z\}_{Z=1}^{4}$, with spectrum $\{-1,1\}$, that is, such that $O_Z:c\rightarrow\{1,-1\}$, for all $c\in Z$. Then, the CHSH parameter \cite{CHSH} can be written as

\be
S\equiv E_{13}+E_{23}+E_{14}-E_{24}.
\ee

\noindent It is well known that, for any set of local correlations, the value of $S$ satisfies $|S|\leq 2$, whereas in the quantum case $|S|\leq 2\sqrt{2}$, the famous Tsirelson bound \cite{tsi_bound}. Moreover, both inequalities can be saturated. 

In \cite{quantum}, it was shown that the maximum value of $|S|$ in macroscopically local theories is also $2\sqrt{2}$. Moreover, any set of correlators $\{E_{13},E_{23},E_{14},E_{24}\}$ arising from a macroscopically local theory admits a quantum representation, and therefore has to satisfy the Tsirelson-Landau-Masanes inequalities \cite{tsirelson,landau,masanes2}

\be
|\sum_{X,Y}\arcsin(E_{XY})-2\arcsin(E_{X'Y'})|\leq \pi,\forall X',Y'.
\ee

The previous results can be easily extended. It can be shown that, for $d=2$ and an arbitrary number of measurement settings $s$, \emph{any} set of two-point correlators $\{E_{XY}\}_{X=1,...,s,Y=s+1,..,2s}$ arising from correlations exhibiting macroscopic locality can also be simulated in a quantum system. See Appendix \ref{corre_proof} for a proof.

\noindent However, not even when $s=d=2$ the set of two-point correlators does contain all the information about the distribution $P(a,b)$. A little thought will convince us that, in order to recover the whole set of probabilities $P(a,b)$, we also need to know the value of the \emph{one-point correlators} $E_Z\equiv \langle O_Z\rangle$. In general, one would have to resort to numerical methods to characterize the set of one-point and two-point correlators compatible with macroscopic locality. In \cite{quantum}, though, it was found that, for the case $s=d=2$, such a set is completely specified by the constraints \footnote{Here we are also implicitly assuming that $\{E_Z,E_{ZT}\}$ arise from a no-signaling set of correlations $P(a,b)$. Note that condition (\ref{necond}) alone does not guarantee that $P(a,b)\geq0$, for all $a,b$.}:

\begin{equation}
\label{necond}
|\sum_{X,Y}\arcsin(\tilde{E}_{XY})-2\arcsin(\tilde{E}_{X'Y'})|\leq \pi,\forall X',Y',
\end{equation}

\noindent where $\tilde{E}_{XY}=(E_{XY}-E_XE_Y)/\sqrt{(1-E_X^2)(1-E_Y^2)}$. From (\ref{necond}), we can see that, although $2\sqrt{2}$ is the maximum violation of the CHSH inequality, there exist macroscopically local distributions that attain this value, but nevertheless are slightly unbiased (i.e., $E_Z\not=0$, for some $Z$). Such distributions, therefore, are not compatible with a quantum theory of nature \cite{wolf}. So even in this simple scenario, we can see that the inclusion $Q\subset Q^1$ is strict, although in this case both sets are extremely close.

\begin{figure}
  \centering
  \includegraphics[width=9.5 cm]{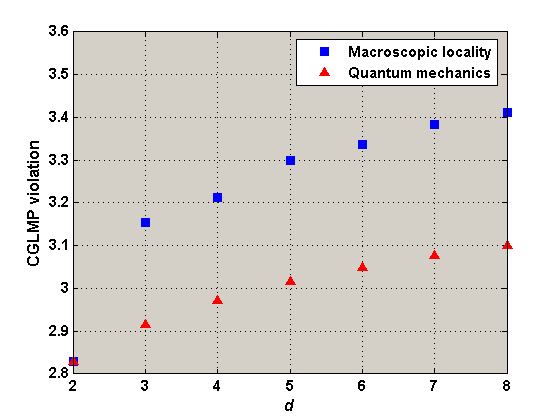}
  \caption{{\bf Departure from Quantum Mechanics.} As the CHSH parameter, the CGLMP parameters \cite{CGLMP} are linear functions $S_d$ on the probabilities $P(a,b)$, to be applied over scenarios with $s=2$ and arbitrary $d$, and, as the CHSH parameter, the maximum possible value of these functions under the assumption of locality is 2. The plot above shows the difference between the maximal possible value of $S_d$ inside $Q$ as well as inside $Q^1$ as a function of $d$ (where the numerical data were taken from \cite{quantum}). At first glance, there seems to be room for a wide variety of macroscopically local theories other than quantum mechanics. This suggests that, if our universe happened to be described by one of these, we might be able to experimentally falsify quantum mechanics with a sufficient number of detectors.}
  \label{detectors}
\end{figure}

The similarities between $Q^1$ and $Q$ decrease, though, as we increase the number of available detectors $d$, as shown in Figure \ref{detectors}. This opens the possibility of disproving quantum mechanics in the future via a Bell-type experiment.

\section{Discussion}

In this article, we have introduced a new physical principle, macroscopic locality, and considered its implications, together with no-signaling, as a fundamental law of nature. We have identified the set $Q^1$ of bipartite correlations that can arise in theories limited by the former two axioms, and commented on its differences and similarities with standard Quantum Mechanics. 

However, nothing has been said about the dynamics of such postquantum theories. Actually, a promising line of research is to try to extend our up-to-bottom approach to restrict the set of microscopic theories that recover General Relativity in some macroscopic limits. This could be of high importance, since it would allow to make model-independent astrophysical predictions, based more on general axioms confirmed by observation than on current theoretical fashions.

It also remains to know how our world would behave if it did admit correlations slightly beyond the set $Q^1$. Would this have negligible experimental consequences, or on the contrary, could we ``distill'' those correlations somehow in order to obtain arbitrary violations of macroscopic locality? And, even if we could not, would the \emph{mere} existence of deviations from macroscopic locality lead to a drastic change in our understanding of the universe?


\section*{Acknowledgements}
The authors thank Martin Plenio for useful discussions. This work is supported by the EU Integrated Project QAP.

\begin{appendix}
\section{$Q^1$ is the set of macroscopically local correlations}
\label{charac_proof}

Using the terminology of \cite{quantum}, $Q^1$ corresponds to the set of all correlations that admit a positive semidefinite \emph{certificate of order 1}. This means that $P(a,b)\in Q^1$ iff there exists a positive semidefinite matrix $\gamma$ of the form

\be
\gamma=\left(\begin{array}{ccc}1&\vec{P}^T_A&\vec{P}^T_B\\\vec{P}_A&\tilde{Q}&\tilde{P}^T\\\vec{P}_B&\tilde{P}&\tilde{R}\end{array}\right),
\label{Q1}
\ee

\noindent where $\vec{P}_A$ ($\vec{P}_B$) is the vector of probabilities $P(a)$ ($P(b)$). Here $\tilde{P}$ is a matrix whose columns are numbered by Alice's outcomes $a$ and whose rows are numbered by Bob's possible outcomes $b$, and such that $\tilde{P}_{ab}=P(a,b)$. The entries of $\tilde{Q}$ (with rows and columns numbered by Alice's measurement outcomes) and $\tilde{R}$ (with rows and columns numbered by Bob's measurement outcomes) are partially determined by the relations

\begin{eqnarray}
\tilde{Q}_{aa'}&=&\delta_{aa'}P(a)\mbox{ if } X(a)=X(a'),\nonumber\\
\tilde{R}_{bb'}&=&\delta_{bb'}P(b)\mbox{ if } Y(b)=Y(b').
\label{q}
\end{eqnarray}

The matrix $\gamma$ is very similar to the matrix $\Gamma$ whose positivity was equivalent to macroscopic locality. The latter (see the main article) was defined as a matrix of the form

\be
\Gamma=\left(\begin{array}{cc}Q&P\\P^T&R\end{array}\right),
\ee

\noindent where $P_{ab}=P(a,b)-P(a)P(b)$, and

\begin{eqnarray}
Q_{a,a'}&=&\delta_{aa'}P(a)-P(a)P(a'),\mbox{ if } X(a)=X(a'),\nonumber\\
R_{b,b'}&=&\delta_{bb'}P(b)-P(b)P(b'),\mbox{ if } Y(b)=Y(b').
\end{eqnarray}

The proof that $Q^1$ is equivalent to the set of macroscopically local correlations follows directly from Schur's theorem \cite{manal}:

\begin{theo}
Let $H$ be a matrix of the form

\be
H=\left(\begin{array}{cc}E&F\\F^T&G\end{array}\right),
\ee

\noindent such that $E>0$. Then, $H\geq 0$ iff $G-F^TE^{-1}F\geq 0$.
\end{theo}

Let us apply this theorem to expression (\ref{Q1}). Since $1>0$, the positivity of $\gamma$ is equivalent to the positivity of 

\be
\left(\begin{array}{cc}\tilde{Q}&\tilde{P}^T\\\tilde{P}&\tilde{G}\end{array}\right)-\left(\begin{array}{c}\vec{P}_A\\ \vec{P}_B\end{array}\right)\cdot(\vec{P}^T_A,\vec{P}^T_B).
\ee

\noindent The reader can check that the last expression is equal to $\Gamma$. To go from one positive semidefinite matrix to the other is thus enough to change the undetermined coefficients of the corresponding matrix according to the rule

\begin{eqnarray}
& &Q_{aa'}=\tilde{Q}_{aa'}-P(a)P(a'),\nonumber\\
& &R_{bb'}=\tilde{R}_{bb'}-P(b)P(b').
\end{eqnarray}

The equivalence between $Q^1$ and the set of all microscopic correlations that give rise to local macroscopic correlations has been proven.

\section{Closure under wiring}
\label{wiring_proof}

When Alice or Bob perform \emph{wiring} over their subsystems, the effective measurement setting $\bar{X}$ of the resulting set of correlations is to be identified with the particular \emph{measurement strategy} they choose at this respect. For example, if Alice and Bob share three systems with $s=d=2$, a measurement strategy for Alice could be as follows: Alice measures her first system with setting 1. If the outcome is $-1$, she measures her second system with setting 2. On the contrary, if the outcome of her first measurement is 1, she measures her third subsystem with setting 1. Finally, the outcome of her second measurement will correspond to the measurement to be implemented in the remaining system. The effective outcome of her virtual box will be the output of the last system measured.

There are two interesting things to point out here:

\begin{enumerate}

\item The order in which the systems are measured is not a priori determined.

\item Alice is not using all the information she has gathered in order to come up with a final outcome. Note that there are several ways in which she can arrive at, say, effective result 1. If Alice wanted to keep all the information received after applying her measurement strategy, she would have to establish a bijective correspondence between her effective outcomes and the physical outcomes of the three subsystems. In short, for each effective measurement $\bar{Z}$, there should be 8 different effective possible outcomes instead of only 2.
\end{enumerate}

If, for whatever measurement strategy, Alice labels the effective outcome by the outcomes of all her measured subsystems, we will say that Alice is performing a \emph{complete} measurement.

To prove that $Q^1$ is closed under wiring, we will need the following trivial lemma.

\begin{lemma}{Closure under identification of outputs}\\
Let $P(a,b)\in Q^1$, and let $Q(a,b)$ be the new set of correlations that arises when Alice and Bob identify several of their measurement outcomes, i.e., when Alice and Bob relabel the possible outcomes of some measurements in such a way that two or more different outcomes will have the same label. Then, $Q(a,b)\in Q^1$.
\end{lemma}

\begin{proof}
Suppose, for simplicity, that Alice has just identified the outcomes $a$ and $a'$ of measurement $X$. If we prove that the resulting distribution $Q(a,b)$ is macroscopically local, then, by induction, we can arrive at any possible identification of outputs of both Alice and Bob. 

Now, let $P(I_a,I_{a'},\vec{I}_{A\backslash \{a,a'\}},\vec{I}_B)$ be a local hidden variable model for the intensities registered by Alice and Bob when they bring $P(a,b)$ to the macroscopic scale. If, under the new wiring, $a$ is identified with $a'$, it is straightforward that the new outcome will give rise to an intensity $I_{\bar{a}}=I_{a}+I_{a'}$, and therefore a hidden variable model for $Q(a,b)$ will be given by $Q(I_{\bar{a}},\vec{I}_{A\backslash \{\bar{a}\}},\vec{I}_B)=\int dI P(I_{\bar{a}}-I,I,\vec{I}_{A\backslash \{a,a'\}},\vec{I}_B)$.
\end{proof}

We are now in a position to prove that $Q^1$ is closed under wiring. Suppose that Alice and Bob share $n$ independent (and, in general, different) physical systems $\{P_i(a,b)\}_{i=1}^n$ such that $P_i\in Q^1$, for all $i=1,...,n$. 

The above Lemma implies that, in order to prove that any wiring of $\{P_i\}_{i=1}^n$ leads to a macroscopically local distribution, it suffices to restrict Alice and Bob to perform strategies where all subsystems are measured and the effective outcomes $\bar{a},\bar{b}$ are given by the vectors of outcomes of the $n$ systems. That is, $\bar{a}=(\bar{a}(1),\bar{a}(2),...,\bar{a}(n)),\bar{b}=(\bar{a}(1),\bar{a}(2),...,\bar{a}(n))$. We can also restrict to deterministic strategies, since any non deterministic strategy can be modeled by adding an extra system that outputs local random symbols according to a given probability distribution.

As we saw in the previous section, if $P_i\in Q^1$, there exists a positive semidefinite matrix $\gamma^i$ of the form (\ref{Q1}). On the other hand, for any positive semidefinite matrix $H$ there exists a set of vectors $\{\vec{v}_k\}_k$ such that $H_{kl}=\vec{v}_k\cdot\vec{v}_l$ \cite{manal}. Perform such a decomposition over the matrices $\{\gamma^i\}$ in order to obtain the vectors $\vec{v}^i_c$, where $c$ can be either the symbol $\id$ (corresponding to $\gamma^i_{11}=1$, for example) or a measurement outcome $a,b$ in system $i$. Define the vector $\vec{w}_\id \equiv \bigotimes_{i=1}^n\vec{v}^i_{\id}$, and, for any outcome $\bar{a}$ ($\bar{b}$) of any strategy $\bar{X}$ ($\bar{Y}$) of Alice's (Bob's), define the vectors $\vec{w}_{\bar{a}}\equiv \bigotimes_{i=1}^n\vec{v}^i_{\bar{a}(i)}$ ($\vec{w}_{\bar{b}}\equiv \bigotimes_{i=1}^n\vec{v}^i_{\bar{b}(i)}$). It is straightforward to see that the matrix $\bar{\gamma}_{\bar{c}\bar{c'}}\equiv \vec{w}_{\bar{c}}\cdot \vec{w}_{\bar{c'}}$ is positive semidefinite. We will show that this matrix is, indeed, a certificate of order 1, i.e., a matrix of the form (\ref{Q1}), for the new set of correlations $P(\bar{a},\bar{b})$.

First, 

\begin{eqnarray}
& &\tilde{P}_{\bar{a}\bar{b}}=\vec{w}_{\bar{a}}\cdot\vec{w}_{\bar{b}}=\prod_{i=1}^n\vec{v}_{\bar{a}(i)}^i\cdot \vec{v}_{\bar{b}(i)}^i=\nonumber\\
& &=\prod_{i=1}^n \gamma^i_{\bar{a}(i)\bar{b}(i)}=\prod_{i=1}^n P(\bar{a}(i),\bar{b}(i))=P(\bar{a},\bar{b}),
\end{eqnarray}

\noindent as it should be. In an analogous way, we can see that the vectors $\vec{P}_{\bar{A}},\vec{P}_{\bar{B}}$ in (\ref{Q1}) are recovered, and that $Q_{\bar{a},\bar{a}}=P(\bar{a})$, $R_{\bar{b},\bar{b}}=P(\bar{b})$. Of course, $\bar{\gamma}_{\id\id}=\prod_{i=1}^n\|\vec{v}^i_\id\|^2=1$.

It only rests to see that $\tilde{Q}_{\bar{a},\bar{a}'}=0$ when $\bar{a}\not=\bar{a}'$ and $\bar{X}(\bar{a})=\bar{X}(\bar{a}')$, and the analogous relation for $\tilde{R}$. 

\noindent Note that, because Alice's strategies are complete, despite the order of the measurements is not a priori known, the ``measurement path'' leading to two different effective outcomes $\bar{a},\bar{a}'$ corresponding to the same measurement strategy $\bar{X}$ is identical from the beginning until one of the physical measurements performed ends up with a different outcome. This means that there exists an $i$  such that $\bar{a}(i)\not=\bar{a}'(i)$ but $X(i)(\bar{a}_i)=X(i)(\bar{a}'_i)$, and so $\vec{v}^i_{\bar{a}(i)}\cdot \vec{v}^i_{\bar{a}'(i)}=0$. This implies that relations (\ref{q}) hold completely.

The new set of correlations $P(\bar{a},\bar{b})$ thus obeys macroscopic locality.

\section{Any set of two-point correlators accessible in $Q^1$ admits a quantum mechanical model}
\label{corre_proof}

Consider an $s,d$ bipartite scenario, with $d\geq 2$, and define a set of observables $\{O_Z\}_{Z=1}^{2s}$, with spectrum $\{-1,1\}$. As before, $E_{XY}$ will denote the two-point correlator of $O_X,O_Y$, i.e., $E_{XY}\equiv\langle O_XO_Y\rangle$.

What we are going to prove next is that, for any set $\{E_{XY}\}_{X,Y}$ arising from $P(a,b)\in Q^1$, there exists a quantum state $\rho\in B(\H_A\otimes\H_B)$ and a set of hermitian operators $\{\tilde{O}_X\in B(\H_A),\tilde{O}_Y\in B(\H_B)\}$, with $\mbox{spec}(\tilde{O}_X,\tilde{O}_Y)=\{-1,1\}$, such that $E_{X,Y}=\tr(\rho \tilde{O}_X\tilde{O}_Y)$.

First, notice that, if there exists a local hidden variable model for all macroscopic variables of the form $\bar{I}_a,\bar{I}_b$, then there exists a local hidden variable model for the variables $\bar{I}_Z\equiv\sum_{c\in Z}O_Z(c)\bar{I}_c$. 

Since they are a linear combination of gaussian variables, these variables are gaussian, with a positive semidefinite covariance matrix $\Gamma$, with $\Gamma_{XY}=\langle I_XI_Y\rangle=\langle O_XO_Y\rangle-\langle O_X\rangle\langle O_Y\rangle$, and $\Gamma_{ZZ}=\langle O_Z^2\rangle-\langle O_Z\rangle^2$. Taking into account that $\langle O_Z^2\rangle=1$, we have that

\be
\left(\begin{array}{cc}F&E\\E^T&G\end{array}\right)\equiv\Gamma+\left(\begin{array}{c}\langle O_1\rangle\\\langle O_2\rangle\\ \vdots\end{array}\right)(\langle O_1\rangle,\langle O_2\rangle,...)\geq \Gamma\geq 0,
\label{corre}
\ee

\noindent where $F_{XX}=G_{YY}=1$ for all indices $X$, $Y$ and $(E)_{XY}=E_{XY}$ are the two-point correlators of $O_X,O_Y$ defined above. The fact that the matrix on the left hand side of equation (\ref{corre}) is positive semidefinite, implies that there exists a quantum mechanical system able to reproduce the correlators $E_{XY}$ \cite{wehner}.

This proof can be extended trivially to deal with the case of two point correlators originated by observables whose spectrum is \emph{in} $[1,-1]$, instead of \emph{being} $\{1,-1\}$.

\end{appendix}


\begin{thebibliography}{000}
\bibitem{polchinski}
Polchinski, J. \emph{String theory}, vols. {\bf 1} and {\bf 2}, (Cambridge University Press, 1998).
\bibitem{rovelli}
Rovelli, C. \emph{Quantum gravity} (Cambridge University Press, 2004); Thiemann, T.
Lectures on loop quantum gravity, \emph{Lecture Notes in Physics, {\bf 541},
2003}; Smolin, L. An invitation to loop quantum gravity, \emph{hep-th/0408048},
(2004).
\bibitem{page}
Page, D. N. Black Hole Information. \emph{Proc. 5$^th$ Canadian Conf. on General Relativity and Relativistic Astrophysics}. (R. Mann and R. McLenaghan, 1994).
\bibitem{popescu}
Rohrlich, D., Popescu, S. Nonlocality as an axiom for quantum theory. \emph{Found. Phys.}, {\bf 24}, 3, 279, (1995).
\bibitem{acin}
Masanes, Ll., Ac\'in, A. and Gisin, N., General properties of Nonsignaling Theories. \emph{Phys. Rev. A.} {\bf 73}, 012112, (2006).
\bibitem{barrett}
Barnum, H., Barrett, J., Leifer, M., Wilce, A. A generalized no-broadcasting theorem. \emph{Phys. Rev. Lett.} {\bf 99}, 240501 (2007).
\bibitem{masanes}
Masanes, Ll. Universally-composable privacy amplification from causality constraints. \emph{Phys. Rev. Lett.} {\bf 102}, 140501 (2009).
\bibitem{brassard}
Brassard, G., Buhrman, H., Linden, N., Methot, A. A., Tapp A., Unger, F., A limit on nonlocality in any world in which communication complexity is not trivial. \emph{Phys. Rev. Lett.}, {\bf 96} 250401, (2006).
\bibitem{scarani}
Pawlowski, M., Paterek, T., Kaszlikowski, D., Scarani, V., Winter, A., and Zukowski, M., A new physical principle: Information Causality. \emph{arXiv:0905.2292}.
\bibitem{scarani2}
Allcock, J., Brunner, N., Pawlowski, M., Scarani, V. Recovering part of the quantum boundary from information causality. \emph{arXiv:0906.3464}.
\bibitem{brunner3}
Bancal, J.-D., Branciard, C., Brunner, N., Gisin, N., Popescu, S., Simon, C. Testing a Bell inequality in multi-pair scenarios. \emph{Phys. Rev. A} {\bf78}, 062110 (2008).
\bibitem{central}
Tijms, H. \emph{Understanding Probability: Chance Rules in Everyday Life} (Cambridge University Press, 2004).
\bibitem{quantum}
Navascu\'es, M., Pironio, S., Ac\'in, A. A convergent hierarchy of semidefinite programs characterizing the set of quantum correlations. \emph{New J. Phys.} {\bf 10}, 073013 (2008).
\bibitem{sdp}
Vandenberghe, L. and Boyd, Semidefinite programming. \emph{S. SIAM Review} {\bf 38}, 49 (1996).
\bibitem{CHSH}
Clauser, J. F., Horne, M. A., Shimony, A. , and Holt, R. A. Proposed Experiment to Test Local Hidden-Variable Theories. \emph{Phys. Rev. Lett.} {\bf 23}, 880 (1969).
\bibitem{tsi_bound}
Cirel'son, B. S. Quantum Generalizations of Bell's Inequality. \emph{Lett. Math. Phys.} {\bf 4}, 93 (1980).
\bibitem{brunner}
Brunner, N., Skrzypczyk, P. Non-locality distillation and post-quantum theories with trivial communication complexity. \emph{Phys. Rev. Lett.} {\bf 102}, 160403 (2009).
\bibitem{tsirelson}
Tsirelson, B. Quantum analogues of the Bell Inequalities. The case of two spatially separated domains. \emph{J. Sov. Math.} {\bf 36}, 557 (1987).
\bibitem{landau}
L. Landau, Empirical two-point correlation functions. \emph{Found. Phys.} {\bf 18}, 449 (1988).
\bibitem{masanes2}
Masanes, Ll. Necessary and sufficient condition for quantum-generated correlations. \emph{quant-ph/0309137}.
\bibitem{wolf}
Werner, R. and Wolf, M. Bell inequalities and entanglement. \emph{QIC}, {\bf1}, 3, 1 (2001).
\bibitem{CGLMP}
Collins, D., Gisin, N., Linden, N., Massar, S., Popescu, S. Bell inequalities for arbitrarily high dimensional systems. \emph{Phys. Rev. Lett.} {\bf88}, 040404 (2002).
\bibitem{manal}
Horn, R. A. and Johnson, C. R. Matrix Analysis \emph{(Cambridge University Press, 1999)}.
\bibitem{wehner}
Wehner, S. Tsirelson bounds for generalized Clauser-Horne-Shimony-Holt inequalities, \emph{Phys. Rev. A}, {\bf73}, 022110 (2006).
\end{thebibliography}
\end{document}